# A new calculation formula of the nuclear cross-section of therapeutic protons

W. Ulmer

Strahlentherapie Nordwürttemberg and Research Group Radiation Physics and Medical Physics, University of Zürich, Zürich Swizzerland

Email: waldemar.ulmer@gmx.net

**Abstract**

We have previously developed for nuclear cross-sections of therapeutic protons a calculation model, which is founded on the collective model as well as a quantum mechanical many particle problem to derive the S matrix and transition probabilities. In this communication, we show that the resonances can be derived by shifted Gaussian functions, whereas the unspecific nuclear interaction compounds can be represented by an error function, which also provides the asymptotic behavior. The energy shifts can be interpreted in terms of necessary domains of energy to excite typical nuclear processes. Thus the necessary formulas referring to previous calculations of nuclear cross-sections will be represented in section 2. The mass number $A_N$ determines the strong interaction range, i.e. $R_{Strong} = 1.2 \cdot 10^{-13} \cdot A_N^{1/3}$ cm. The threshold energy $E_{Th}$ of the energy barrier is determined by the condition $E_{strong} = E_{Coulomb}$. A linear combination of Gaussians, which contain additional energy shifts, and an error function incorporate a possible representation of Fermi-Dirac statistics, which is applied here to nuclear excitations and reaction with release of secondary particles. The new calculation formula provides a better understanding of different types of resonances occurring in nuclear interactions with protons. The present study is mainly a continuation of the papers [1 - 3].

1. Introduction

The knowledge of the total nuclear cross-section $Q^{tot}$ of protons is an important impact with regard to sophisticated features of therapy planning, since $Q^{tot}$ provides essential information of the following aspects: Decrease of the fluence of primary protons $\Phi_{pp}$ and release of secondary particles and their transport (secondary protons, neutrons, clusters like $H_2^1$, $H_3^1$, $He_3^2$, $He_4^3$, heavy recoil nuclei, which usually undergo either a ß$^+$ or ß$^-$ decay with additional emission of a γ-quant). With regard to secondary protons with have to differ between two kinds, namely protons resulting from nuclear reactions with production of heavy recoils and those protons, which are, in reality, primary protons and have undergone elastic and inelastic scatter by strong interactions according to the Breit-Wigner formula. Elastic scatter by nuclear forces is only a deflection of the projectile protons with additional energy-momentum transfer to the whole target nucleus, whereas inelastic scatter is connected to excitations of nuclear vibrations, rotations and transitions to excited states without releasing other nuclear particles, i.e. some quantum number will be changed. The resonances due to Breit-Wigner formula represent the main part of the resonance domain $E_{res}$ according to Figure 1. Nuclear reaction types cannot be regarded as simple resonances, they mainly occur for proton energies $E > E_{res}$.

In previous publications [2, 3], we have developed a calculation method based on a nonlinear and nonlocal Schrödinger equation with a Gaussian kernel and on an interacting many-body system containing strong interactions, spin-orbit coupling and Coulomb interaction with inclusion of various excited configurations. The results of these calculation methods, which provide excited nuclear states, virtual compounds and nuclear reactions via S-matrix, transition probabilities and finally, total nuclear cross-sections $Q^{tot}(E)$, can be translated to the collective nuclear excitation model. This model only uses Z and $A_N$ as parameters and suitable analytic functions to describe all properties of $Q^{tot}(E)$. The complete contents of this figure have been previously discussed. For protons, a threshold energy $E_{Th}$ exists to surmount the potential barrier of the oxygen nucleus. Thus for proton energies less than $E_{Th}$ nuclear reactions cannot occur. At $E = E_{res} = 20.12$ MeV $Q^{tot}(E)$ shows a maximum value, and thereafter, it decreases exponentially to reach the asymptotic behavior at about $E = 100$ MeV. According to an integration procedure [2, 6] the analytic version of Figure 1 (equation (1)) provides the decrease of primary protons. The following integration procedure has to be carried out:

$$\int_{\Phi_0}^{\Phi} d\Phi / \Phi \;=\; \ln(\Phi / \Phi_0) \;=\; Z \cdot \rho \cdot (N_{Avogadro} / A_N) \cdot \int_{E_0}^{E} Q^{tot}(E) \cdot dE \cdot [dE/dz]^{-1} \}. \quad (1)$$

By that, we obtain the following formula for the fluence decrease of protons:

$$\Phi_{pp} = \Phi_0 \cdot \tfrac{1}{2} \cdot [\, erf((R_{csda} - z))/\tau) + 1\,][1 - ((E_0 - E_{Th})/Mc^2)^{1.032} \cdot z/R_{csda}\,]. \quad (2)$$

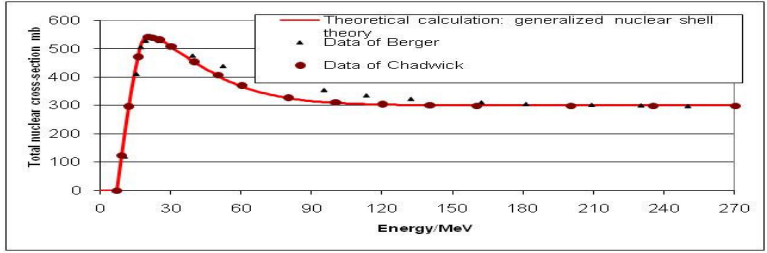

**Figure 1**: Total nuclear cross-section of the proton – oxygen interaction [4, 5].

With regard to oxygen we have to put $E_{Th}$ = 7 MeV (O) and $Mc^2$ = 938.27 MeV. Equation (2) can be summarized by Figure 2.

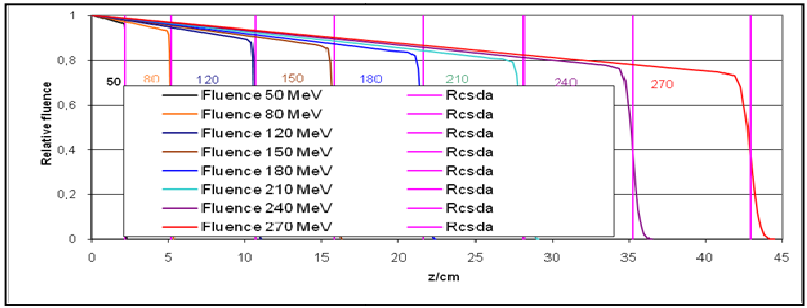

**Figure 2:** Decrease of the fluence of primary protons in water.

A further result is that in the environment of a nucleus the effective potential can be calculated by a linear combination of two Gaussians. This is not true for longer ranges of Coulomb forces. However, this is not a simple task, since the shielding of the nuclear repulsion by the shell electrons has to be accounted for. The nuclear potential according to Figure 3 and the related parameters can be best calculated by equation (3), which assumes the shape:

$$\left. \begin{array}{l} \varphi(r) = V_0 \cdot \exp(-r^2/\sigma_0^2) + V_1 \cdot \exp(-r^2/\sigma_1^2) \\ V_0 = -27.7592 \; MeV \; ; V_1 = 7.75935 \; MeV \; ; \sigma_0 = 0.423901 \; ; \sigma_1 = 3.37402 \end{array} \right\} .(3)$$

The units of this formula are stated in terms of units of $R_{strong}$ (range of strong interaction) according to Figure 3.

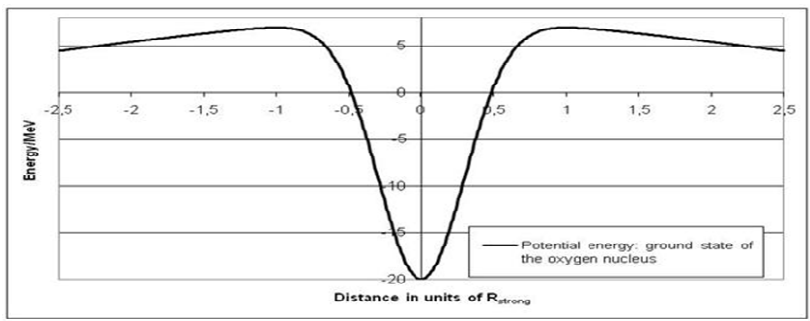

**Figure 3:** Nuclear potential energy of oxygen.

## 2. Methods

In previous publications [2, 3] we have performed the task to calculate all nuclear properties to determine finally with the help of the S-matrix and transition probability of all possible configuration states the total nuclear cross-section $Q^{tot}(E)$ in terms of Z and $A_N$. The results are closely related to the Bethe-Weizsäcker formula for the nuclear binding energy. The old formula for the calculation of $E_{Th}$ and $Q^{tot}(E)$ according to a previous paper will be stated in section 2.1. In order to determine all these properties we have to know the threshold energy $E_{Th}$, i.e. that energy necessary to surmount the nuclear potential mount according to Figure 3. There exists also the possibility to circumvent this threshold restriction by the quantum mechanical tunneling process with energy E <

$E_{Th}$, but this is only a rather small effect, which leads to a little roundness of $Q^{tot}(E)$ at $E < E_{Th}$. This aspect will be considered in section 2.3.

## 2.1. Summary of previous investigations

According to the results presented in [2, 3] we need for the calculation of $Q^{tot}(E)$, at first, the threshold energy $E_{Th}$ as a function of Z and $A_N$. This function can be obtained via the formulas (4 - 6). The second step provides the determination of the total nuclear cross-section $Q^{tot}(E)$, which can be accounted for by preceding studies [2, 6] with the help of equations (7 – 9). Thus the formulas (4 – 6) necessary to determine the threshold energy $E_{Th}$ by a balance equation of Coulomb repulsion and oscillator model for strong interactions.

$$F = (A_N/2Z) \cdot$$
$$\cdot (a_0 + a_1/A_N + a_2/A_N^{p1} + a_3/A_N^{p2} + a_4/A_N^{p3}) \quad (4)$$

$$E_{Th} = C_F \cdot Z^\kappa \cdot F(Z, A_N); \kappa = 1.659 \ (Z \geq 6)$$
$$E_{Th} = C_F \cdot Z^\kappa \cdot F(C_6^{12}) \cdot (6-Z) \ (Z < 6) \quad (5)$$

$$\kappa = 1.659 + 0.341 \cdot (6-Z)/5 \quad (6)$$

$$A_{Th} = \exp(-(E_{Th} - E_{res})^2/\sigma_{res}^2)$$
$$E_m = E_{res} - E_{Th}; \sigma_{res} = \sqrt{\pi} \cdot E_m \quad (7)$$

In the domain of the resonance energy $E_{res}$ we obtain:

$$Q^{tot} = Q^{tot}_{max} \cdot [\exp(-(E-E_{res})^2/\sigma_{res}^2) - A_{Th}] \cdot (1-A_{Th})^{-1} \quad (if \ E_{Th} \leq E_{res})$$
$$Q^{tot} = Q^{tot}_{max} \cdot \exp(-(E-E_{res})^2/2\sigma_{res}^2) \ (if \ E_{res} < E < E_c); E_c = E_{res} + \sqrt{-2 \cdot \ln(I_c)} \quad (8)$$
$$Q^{tot} = Q^{tot}_c - (Q^{tot}_c - Q^{tot}_{as}) \cdot \tanh[(E-E_c)/\sigma_{as}] \ (if \ E > E_c) \ Q^{tot}_c = I_c \cdot Q^{tot}_{max}; \sigma_{as} = \sigma_{res} \cdot (Q^{tot}_c - Q^{tot}_{as})/(Q^{tot}_{max} \cdot \sqrt{-2 \cdot \ln(I_c)}) \quad (9)$$

Table 1: Parameters for the evaluation of equations (7 – 9).

| Nucleus | $E_{Th}$/MeV | $E_{res}$/MeV | $\sigma_{res}$/MeV | $\sigma_{as}$/MeV | $Q^{tot}_{max}$/mb | $Q^{tot}_c$/mb | $Q^{tot}_{as}$/mb |
|---|---|---|---|---|---|---|---|
| C | 5.7433 | 17.5033 | 21.1985 | 27.1703 | 447.86 | 426.91 | 247.64 |
| O | 6.9999 | 20.1202 | 23.2546 | 34.1357 | 541.06 | 517.31 | 299.79 |
| Ca | 7.7096 | 25.2128 | 35.6329 | 58.4172 | 984.86 | 954.82 | 552.56 |
| Cu | 8.2911 | 33.4733 | 47.6475 | 93.2700 | 1341.94 | 1308.07 | 752.03 |

In the domain of the transition to the asymptotic behavior of the total nuclear cross-section the following formula is applicable:

$$Q^{tot} = Q^{tot}_c - (Q^{tot}_c - Q^{tot}_{as}) \cdot \tanh[(E - E_c)/\sigma_{as}] \ (if \ E > E_c). \quad (9)$$

$$Q^{tot}_c = I_c \cdot Q^{tot}_{max}; \sigma_{as} = \sigma_{res} \cdot (Q^{tot}_c - Q^{tot}_{as})/(Q^{tot}_{max} \cdot \sqrt{-2 \cdot \ln(I_c)}). \quad (9a)$$

In order to give a qualitative motivation of the new calculation method we consider again Figure 1. At first, we have to repeat that $Q^{tot}(E)$ is not only restricted to proper nuclear reactions of protons with release of secondary particles: Thus for proton energies $E > E_{Th}$ (the threshold energy $E_{Th}$ amounts to 7 MeV) we can verify a rapid increase of $Q^{tot}$ up to a resonance maximum $E_{res} = 20.12$ MeV. This behavior up to the environment of the resonance maximum can be described by a proper Gaussian distribution. What happens in this domain? Elastic scatter of proton at the nuclear potential is dominated by strong interaction and mediated mainly by mesons, if the quantum state of the nucleus is not changed. This implies that in order to satisfy energy-momentum relation only the whole nuclear adopts energy and momentum (kinetic energy), the impinging proton is slightly deflected. If the nucleus is also excited by vibrations, rotations or transitions to an excited configuration, then the whole process is inelastic and by emitting γ-quanta it is damped to finally return to the ground state. These effects are mainly described by the Breit-Wigner formula and its generalization [6 - 8]. It has to be pointed out that the secondary proton under these conditions is still the primary proton, which is deflected by a slightly higher scatter angle compared to Molière multiple scatter theory [9].

## 2.2. New calculation formulas for $Q^{tot}(E)$ and $E_{Th}$

Figure 1 presents the total nuclear cross-section of oxygen; we can verify that, besides the Gaussian distribution of the environment of the maximal value $E_{max}$, after a slower decrease of $Q^{tot}$, the asymptotic behavior is reached (this is certainly valid for proton energies E < 300 MeV). The asymptotic branch can be represented by a suitable error function erf(E), which has to satisfy some further boundary conditions. This error function erf(E) results by an integration over Gaussian resonance distributions of the energy within finite boundaries. The transition from the domain $E_{max}$ to the asymptotic domain can either be represented by a sum of exponential functions or by a Gaussian distribution with an additional energy shift. In every case, this consideration indicates that there exists an alternative representation of the total nuclear cross-section $Q^{tot}(E)$ besides the previously studied one according to equations (7 – 9). Thus equation (10) provides the new formula for $Q^{tot}(E)$, which appears to lead to a better access to quantum mechanical resonance mechanisms expressed by harmonic oscillators. Equation (11) provides all terms necessary for the determination of boundary conditions of the whole problem, and equation (12) an alternative calculation procedure for $E_{Th}$. $A_{boundary}$ has the purpose to ensure that $Q^{tot}(E)$ assumes zero at $E = E_{Th}$, since the Gaussians do not vanish at this position. A modification will be accounted for, when we shall include the tunneling effect.

$$Q^{tot} = w_0 \cdot Q^{tot}_{as} \cdot erf(\tfrac{E-E_{Th}}{\sigma_{as}}) + w_{g0} \cdot \exp(-(\tfrac{E+\delta-E_{res}}{\sigma_{res}})^2) + A_{boundary} + w_{g1} \cdot A_m \cdot \exp(-(\tfrac{E-E_{res}}{\sigma_1})^2) \; w_{g1} \cdot (1-A_m) \cdot \exp(-(\tfrac{E-\gamma-E_{res}}{\sigma_2})^2) \} \quad (10)$$

$$\left. \begin{array}{l} A_{boundary} = (1-A_m) \cdot wg_0 - A_m \cdot wg_{00} + (1-A_m) \cdot (wg_0 - wg_{f_0}) \cdot erf(\tfrac{E-E_{Th}}{\sigma_{as}}) \\ + A_m \cdot erf(\tfrac{E-E_{Th}}{\sigma_{as}}) \cdot (wg_{00} - wg_{f_1}) + wg_{f_3} \cdot w_{Gauss} + w_{Gauss} \cdot wg_0 - A_m \cdot wg_{00} + (1-A_m) \cdot (wg_0 - wg_{f_0}) \cdot erf(\tfrac{E-E_{Th}}{\sigma_{as}}) \\ + A_m \cdot erf(\tfrac{E-E_{Th}}{\sigma_{as}}) \cdot (wg_{00} - wg_{f_1}) + wg_{f_3} \cdot w_{Gauss} + w_{Gauss} \cdot (wg_{f_1} - wg_{f_2}) \\ wg_0 = \exp(-(\tfrac{E_{Th}-\gamma-E_{res}}{\sigma_2})^2); \; wg_{00} = \exp(-(\tfrac{E_{Th}-E_{res}}{\sigma_1})^2); \; wg_{f_0} = \exp(-(\tfrac{E_{res}-E_f-\gamma}{\sigma_2})^2); \; E_f = 270\,MeV \\ wg_{f_1} = \exp(-(\tfrac{E_{res}-E_f}{\sigma_1})^2); \; wg_{f_2} = \exp(-(\tfrac{E_{res}+\delta-E_f}{\sigma_{res}})^2); \; wg_{f_3} = \exp(-(\tfrac{E_{Th}-E_{res}+\delta}{\sigma_{res}})^2) \end{array} \right\} . \quad (10a)$$

Table 2 presents the corresponding parameter values to perform calculations of equation (10) with the help of equation (11). All necessary parameters of equations (10, 11) can be determined via Table 2 and the formula:
$$P_p = C_p \cdot Z^p / A_N^q . \qquad (11)$$

**Table 2:** Parameters of the new cross-section formula (10) and equations (11) using equation (10a).

| Parameters $P_p$ of formula (12) | $C_p$ | p | q |
|---|---|---|---|
| $w_{Gauss}$ | 36.05 | 1.421 | 1.811 |
| δ | 0.09335 | -1.621 | -0.405 |
| γ | -9.155 | 2.396 | 1.763 |
| $\sigma_{res}$ | 0.925 | -1.232 | -1.595 |
| $\sigma_1$ | 17.215 | 0.6375 | 0.31 |
| $\sigma_2$ | 11.575 | 1.13 | 0.38 |
| $\sigma_{as}$ | 1.074 | 1.745 | 2.102 |
| $A_m$ | 0.06257 | -1.102 | -1.335 |
| wg | -4.411 | 25.8712 | -3.302 |

The total nuclear cross-section requires also the threshold energy $E_{Th}$, which can be calculated by the following formula (12), which is easier to handle than formulas (4 – 6):

$$E_{Th} = C \cdot Z^{p} / A_{N}^{q} + D \cdot Z^{p1} / A_{N}^{q1}. \quad (12)$$

The parameters of formula (12) are: C = 6.565304 MeV, p = -0.10368 and q = 0.00481, D = -1.2889, p1 = -0.6597, q1 = -0.6601.

Table 3: Calculation of the threshold energy $E_{Th}$ according to formula 4 – 6 (B) and present formula 12 (A).

| Isotope | A $E_{Th}$/MeV | B $E_{Th}$/MeV |
|---|---|---|
| C 12 | 6.480 | 6.7 |
| C 13 | 6.6100 | 6.61 |
| C 14 | 6.5701 | 6.51 |
| O 16 | 6.9235 | 6.99 |
| Ca 40.06 | 7.86 | 7.75 |
| Cu 63.456 | 8.24 | 8.24 |
| Zn 65.39 | 8.2796 | 8.29 |
| Cs 136 | 8.949 | 9 |
| Cs 137 | 8.9464 | 8.92 |

### 2.3. Calculation procedure of the energy levels of excited states and quantum mechanical tunneling effect for E < $E_{Th}$

The investigations referring to the tunneling effect are as old as the quantum mechanics itself, since this theory has been used to explain the α-decay of heavy nuclei by Gamow. In many textbooks of quantum mechanics the tunnel effect is treated in detail of a particle passing with energy E through a potential box V with E < V . In the mean time, some other important tunnel effects have been studied, e.g. Josephson junctions and, recently, the H bonds between the complementary DNA strands through a double minimum potential [14, and references therein]. After this digression, we intend to return to Figure 3, which refers to the nuclear potential of $O_8^{16}$, but all formulas developed here can be applied to other nuclei, which exhibit similar properties of their nuclear potential. In following, we need the deflection point ξ of this figure. If we reduce formula (3) to one Gaussian with $V_1 = 0$, this point is easy to determine, since vanishing of the second derivation provides $\xi^2 = \sigma_0^2/2$. The deflection point of the nuclear potential according to formula (3) can only be obtained by an iteration procedure:

$$\tfrac{d^2}{dx^2}\varphi(r) = \tfrac{d^2}{dx^2}[V_0 \cdot \exp(-r^2/\sigma_0^2) + V_1 \cdot \exp(-r^2/\sigma_1^2)]. \quad (13)$$

In a first step, we restrict equation (13) to first-order terms; by that, we obtain a quadratic equation in terms of $\xi_0^2$:

$$\xi_0^4 \cdot \tfrac{2V_0}{\sigma_0^4} \cdot (\tfrac{1}{\sigma_0^2} - \tfrac{1}{\sigma_1^2}) + \xi_0^2 \cdot (2 \cdot V_1/\sigma_1^4 + V_1/(\sigma_0^2 \cdot \sigma_1^2) - 3 \cdot V_0/\sigma_0^4) + V_0/\sigma_0^2 - V_1/\sigma_1^2 = 0. \quad (14)$$

In the next step, we have to solve a quadratic equation, too, since we insert the solution with $\xi_0$ into the Gaussians of equation (13) and determine $\xi_1^2$ according this equation by putting $\xi_1^2 = x^2$, i.e. the following step is given by:

$$\left. \begin{array}{l} \xi_1^2 \cdot [-V_0 \cdot \exp(-\xi_0^2/\sigma_0^2) \cdot \tfrac{4}{\sigma_0^4} + V_1 \cdot \exp(-\xi_0^2/\sigma_1^2) \cdot \tfrac{4}{\sigma_1^4}] + \\ + 2 \cdot V_0 \cdot \exp(-\xi_0^2/\sigma_0^2)/\sigma_0^2 - 2 \cdot V_1 \cdot \exp(-\xi_0^2/\sigma_1^2)/\sigma_1^2 = 0 \\ \xi_{n+1}^2 \cdot [-V_0 \cdot \exp(-\xi_n^2/\sigma_0^2) \cdot \tfrac{4}{\sigma_0^4} + V_1 \cdot \exp(-\xi_n^2/\sigma_1^2) \cdot \tfrac{4}{\sigma_1^4}] + \\ + 2 \cdot V_0 \cdot \exp(-\xi_n^2/\sigma_0^2)/\sigma_0^2 - 2 \cdot V_1 \cdot \exp(-\xi_n^2/\sigma_1^2)/\sigma_1^2 = 0 \end{array} \right\} . \quad (14\,a)$$

In principle, this iterative procedure can be repeated, but stopping after step 2 is very sufficient. We now determine the energy levels of the potential type like that of formula (3). It offers to approximate the potential by a 3D harmonic oscillator. However, this is not sufficient, since the energy levels of the excited states are not equidistant [2]. Therefore we use a nonlinear and nonlocal Schrödinger equation, which incorporates nuclear interactions as a self-interacting field. With the help of equation 3 this generalized Schrödinger equations assumes the form:

$$\left. \begin{array}{l} K(\vec{r} - \vec{u}) = V_0 \cdot \exp(-(\vec{r} - \vec{r}\,')^2/\sigma_0^2) + V_1 \cdot \exp(-(\vec{r} - \vec{r}\,')^2/\sigma_1^2) \\ \tfrac{\hbar^2}{2\mu} \cdot \Delta \cdot \Psi + E \cdot \Psi = \lambda \cdot \int d^3u \cdot K(\vec{r} - \vec{r}\,') \cdot |\psi(\vec{r}\,')|^2 \cdot \Psi \end{array} \right\} . (15)$$

The coupling constant λ has to be chosen such that the dimension agrees on both sides, but it can be put λ = 1; the magnetic interaction, i.e. spin-orbit-coupling, can be added using principles previously given [2, 3]. With the help of the generating functions of Hermite polynomials we are able to write equation (15) in the form:

$$\begin{aligned}
E \cdot \psi + \tfrac{\hbar^2}{2M} \cdot \Delta \psi &= \varphi(x,y,z) \cdot \psi = \sum_{n1=0}^{\infty} \sum_{n2=0}^{\infty} \sum_{n3=0}^{\infty} \Phi_{n1,n2,n3} \cdot x^{n1} \cdot y^{n2} \cdot z^{n3} \cdot \psi \\
\Phi_{n1,n2,n3} &= -V_0 \cdot \tfrac{1}{n1!} \cdot \tfrac{1}{n2!} \cdot \tfrac{1}{n3!} \cdot \tfrac{1}{\sigma_0^{n1+n2+n3}} \cdot \int |\psi(x',y',z')|^2 \cdot \exp(-(x'^2+y'^2+z'^2)/\sigma_0^2) \cdot \\
&\quad \cdot H_{n1}(x'/\sigma_0) \cdot H_{n2}(y'/\sigma_0) \cdot H_{n3}(z'/\sigma_0) dx' dy' dz' \\
&\quad + V_1 \cdot \tfrac{1}{n1!} \cdot \tfrac{1}{n2!} \cdot \tfrac{1}{n3!} \cdot \tfrac{1}{\sigma_1^{n1+n2+n3}} \cdot \int |\psi(x',y',z')|^2 \cdot \exp(-(x'^2+y'^2+z'^2)/\sigma_1^2) \cdot H_{n1}(x'/\sigma_1) \cdot H_{n2}(y'/\sigma_1) \cdot H_{n3}(z'/\sigma_1) dx'^3
\end{aligned} \quad (16)$$

The equation above represents a highly inharmonic oscillator equation of a self-interacting field. Since the square of the wave-function is always positive definite, all terms with odd numbers of n1, n2, and n3 vanish due to the anti-symmetric properties of those Hermite polynomials. For $r_c \leq \xi$ (domain with positive curvature), the whole equation is reduced to a harmonic oscillator with self-interaction; the higher-order terms are small perturbations:

$$E \cdot \psi + \tfrac{\hbar^2}{2M} \cdot \Delta \psi = \varphi(x,y,z) \cdot \psi = [\Phi_{0,0,2}(x^2+y^2+z^2) + \Phi_{0,0,0}] \cdot \psi. \quad (17)$$

The solutions of this equation are those of a 3D harmonic oscillator; the classification of the states by $SU_3$ and all previously developed statements with regard to the angular momentum are still valid. The only difference is that the energy levels are not equidistant; this property can easily be verified in one dimension. The usual ground state energy is $\hbar\omega_0/2$. This energy level is lowered by the term $\sim \Phi_{0,0,0}$, depending on the ground-state wave-function. The energy difference between the ground and the first excited state amounts to $\hbar\omega_0$; this is not true in the case above, since the energy levels of all excited states depend on the corresponding eigen-functions themselves (these are still the oscillator eigen-functions!). Next, we will include the terms of the next order, which are of the form $\sim \lambda \cdot (\Phi_{0,2,2}, \Phi_{2,2,0}, \Phi_{2,0,2})$:

$$\begin{aligned}
E \cdot \psi + \tfrac{\hbar^2}{2M} \cdot \Delta \psi &= \varphi(x,y,z) \cdot \psi = [\Phi_{0,0,0} + \Phi_{0,0,2}(x^2+y^2+z^2) + T] \cdot \psi \\
T &= \Phi_{2,2,0} \cdot x^2 \cdot y^2 + \Phi_{2,0,2} \cdot x^2 \cdot z^2 + \Phi_{0,2,2} \cdot y^2 \cdot z^2
\end{aligned} \quad (18)$$

The additional term T represents tensor forces. The whole problem is still exact soluble. In further extensions of the nonlinear/nonlocal Schrödinger equation, we are able to account for spin, isotopic spin, and spin-orbit coupling. The spin-orbit coupling, as an effect of an internal field with nonlocal self-interaction, is plausible, since the extended nucleonic particle has internal structure; consequently, we have to add $H_{so}$ to the nonlinear term:

$$H_{so} \cdot \psi = g_\tau \cdot \tfrac{\hbar \cdot \vec{\sigma}}{4 \cdot M \cdot c^2} \cdot \nabla \varphi \times p \cdot \psi. \quad (19)$$

Ψ is now (at least) a Pauli spinor (i.e. a two-component wave-function), and due to the term $H_{so}$ the $SU_3$ symmetry is broken. We should like to point out that the operation grad φ acts on the Gaussian kernel K:

$$\begin{aligned}
\nabla \varphi &= -\tfrac{2}{\sigma_0^2} \cdot [H_1((x-x')/\sigma_0), H_1((y-y')/\sigma_0), H_1((z-z')/\sigma_0)] \cdot \varphi \\
&\quad - \tfrac{2}{\sigma_1^2} \cdot [H_1((x-x')/\sigma_1), H_1((y-y')/\sigma_1), H_1((z-z')/\sigma_1)] \cdot \varphi
\end{aligned} \quad (20)$$

The expression in the bracket of the previous equation represents a vector, and **p** (**p** → -iℏgrad) acts on the wave-function. Since the neutron is not a charged particle, the spin-orbit coupling of a neutron can only involve the angular momentum of a proton. In nuclear physics, these nonlinear fields are adequate for the analysis of clusters (deuteron, He, etc.). The complete wave-function $\Psi_c$ is now given by the product of a function in configuration space Ψ multiplied with the total spin and isotopic-spin functions.

A further access to solve the above nonlinear/nonlocal equation is obtained by the Ritz' variation principle applied to energy minimum. For this purpose, we start with the following set of functions:

$$\Psi = \sum_{j=0,k=0,l=0}^{N} c_{j,k,l} \cdot [(x/\sigma_0)^j \cdot (y/\sigma_0)^k \cdot (z/\sigma_0)^l \cdot \exp(-r^2/\sigma_0^2)] + c'_{j,k,l} \cdot [(x/\sigma_1)^j \cdot (y/\sigma_1)^k \cdot (z/\sigma_1)^l \cdot \exp(-r^2/\sigma_1^2)]. \quad (21)$$

The expectation value of the energy is given by:

$$\int \Psi^* E \Psi d^3x = \int \Psi^* (-\tfrac{\hbar^2}{2\mu} \Delta) \Psi d^3x + \int \Psi^* H_{so} \Psi d^3x + \int \Psi^*(\vec{r}) \int K(\vec{r}-\vec{r}') |\Psi(\vec{r}')|^2 d^3x' \Psi(\vec{r}) d^3x. \quad (22)$$

It is now the task to determine the coefficients $c_{j,k,l}$ and $c'_{j,k,l}$ in such a way that the energy assumes a minimum value, i.e. the following conditions holds:

$$\delta E(c_{j,k,l}, c'_{j,k,l}) = 0 \Rightarrow \delta c_{j,k,l} = 0; \quad \delta c'_{j,k,l} = 0 \quad (j,k,l = 0,..., N). \quad (23)$$

With the help of the function set (21) all integrals can be evaluated analytically. By that, the task remains to determine the coefficients of the function set by an iterative procedure, since equation (23) yields a third order equation of all coefficients forming the wave-function. However, this so-called self-consistent field method is very familiar in many-particle problems, and, therefore, a detailed description is superfluous. The performance of the above task provides, besides the ground state energy, excited states for $E < E_{Th}$ and virtually excited states, if $E > E_{Th}$. However, we are interested in the role of excited states of a nucleus, when a proton can penetrate the potential wall for $E < E_{Th}$ and virtually excited states are only important with regard to nuclear reaction. Therefore we now turn our interest to the tunneling effect of external protons in connection with potential types like Figure 3.

The quantum mechanical tunneling effect must be treated in three dimensions. We denote by μ the reduced mass 'proton – nucleus', then the Schrödinger equation of the problem reads:

$$\left. \begin{array}{l} H \cdot \Psi = -\frac{\hbar^2}{2\mu} \cdot \Delta\Psi(x,y,z) + \varphi(r) \cdot \Psi(z) = E \cdot \Psi(x,y,z) \\ H_0 = -\frac{\hbar^2}{2\mu} \cdot \Delta, \quad H_1 = \varphi(r), \quad H = H_0 + H_1 \\ \varphi(r) = -V_0 \cdot \exp(-r^2/\sigma_0^2) + V_1 \cdot \exp(-r^2/\sigma_1^2) \end{array} \right\} . (24)$$

The boundary condition for the proton energy is $0 < E < E_{Th}$. Since a closed analytical solution is unknown, we shall make use here of the Dyson series; this is similar to the Feynman propagator method, and easy to manipulate for plain waves and Gaussians (kernels, potentials, etc.). In absence of external perturbation, the time evolution operator reads:

$$U(t, t_0) = \exp(-i H_0 (t - t_0)/\hbar). \quad (25)$$

Using the interaction picture of quantum mechanics and the definition of $U(t, t_0)$ according to equation (25), the interaction term can be transformed to the expression:

$$H_I(t_0, t) = U^{-1} \cdot H_1 \cdot U \quad (26)$$

Based on the interaction picture, the resolution operator $U_I(t_0, t)$ is given by the integral equation:

$$U_I(t_0, t) = 1 - \frac{i}{\hbar} \int_{t_0}^{t} H_I(t_0, t') \cdot U_I(t_0, t') dt'. \quad (27)$$

The Dyson expansion of equation (27) provides the following expression:

$$U_I(t_0, t) = 1 - \frac{i}{\hbar} \int_{t_0}^{t} H_I(t_0, t') dt' + (\frac{-i}{\hbar})^2 \cdot \int_{t_0}^{t}\int_{t_0}^{t'} H_I(t_0, t'') \cdot H_I(t_0, t') \cdot dt'' dt' + .....$$

$$+ (\frac{-i}{\hbar})^n \cdot \int_{t_0}^{t}\int_{t_0}^{t'}.......\int_{t_0}^{t^{(n-1)}} H_I(t_0, t^{(n-1)}) \cdots H_I(t_0, t'') \cdot H_I(t_0, t') \cdot dt^{(n-1)}...dt'' dt'. (28)$$

Within the restriction of our task, namely that $H_0$ is the free-particle operator and $H_1$ a linear combination of 2 Gaussian functions (see equation (25)), the expansion (28) is easy to solve: The unitary operator U related to $H_0$ applied to the terms of $H_1$ again provides Gaussian terms. Since the operator U also acts on plain waves, too, the similar behavior is valid. Thus the following expressions have to subject to iterations (the factor α has to associate with the proper dimension):

$$\exp(\alpha \cdot \Delta) \cdot \frac{1}{(\sqrt{2\pi})^3} \cdot \exp(-i\vec{k} \cdot \vec{x}) = \frac{1}{(\sqrt{2\pi})^3} \cdot \exp(-i\vec{k} \cdot \vec{x}) \cdot \exp(-\alpha \cdot \vec{k}^2). \quad (29)$$

$$\int_{-\infty}^{\infty} \exp(-i \cdot \vec{k} \cdot (\vec{x} - \vec{x}')) \cdot \exp(-\alpha \cdot \vec{k}^2) d^3k = \sqrt{\frac{\pi^3}{\alpha^3}} \cdot \exp(-(\vec{x} - \vec{x}')^2 / 4\alpha). \quad (29a)$$

With respect to the Dyson expansion the terms according to equations (29, 29a) are very convenient, since iterations always yield terms of the same structure. Please note that the kinetic energy term of the proton assumes complex values, if the impinging energy is lower than the potential wall, i.e. the position probability within the positive, repulsive potential suffers damping. Some consequences will be given in the result section.

## 3. Results

The following figures 4 – 7 deal with the nuclei of carbon, oxygen, calcium and copper. They show that both calculation formulas provide equivalent results; we use the following abbreviations: FM (former method) and PM (present method). However, the chosen type of functions used in this communication offers new ways to analyze nuclear cross-sections due to the advantages of Gaussian functions and kernels.

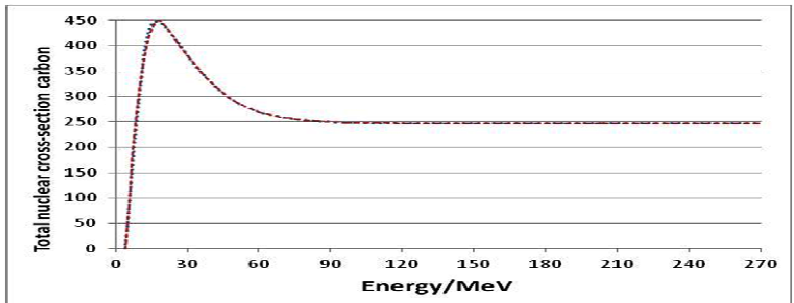

**Figure 4:** Nuclear cross-section of the interaction proton – carbon nucleus; solid line: PM, dots: FM.

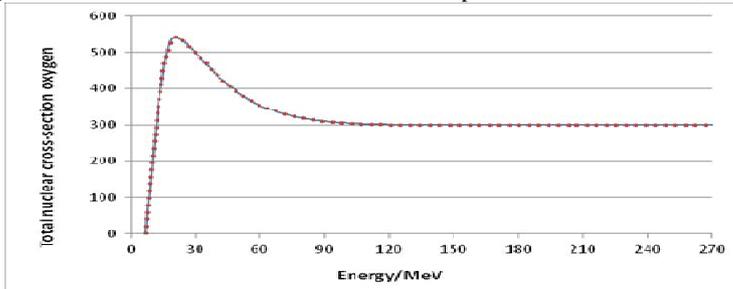

**Figure 5:** Nuclear cross-section of the interaction proton – oxygen nucleus; solid line: PM, dots: FM.

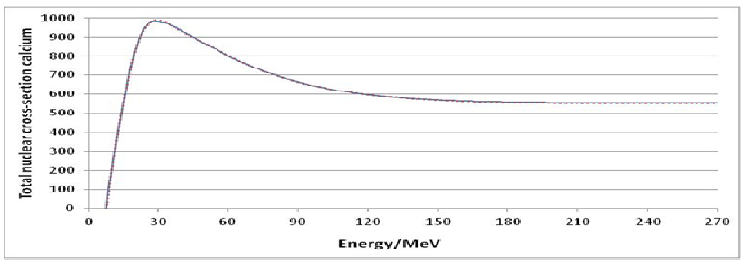

**Figure 6:** Nuclear cross-section of the interaction proton – calcium nucleus; solid line: PM, dots: FM.

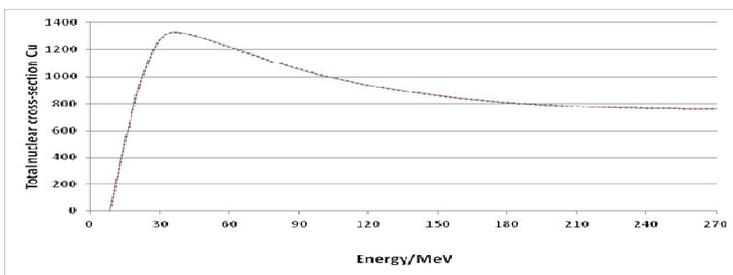

**Figure 7:** Nuclear cross-section of the interaction proton – copper nucleus; solid line: PM, dots: FM.

The following Figures 8 – 11 show with regard to the most important nuclei in radiotherapy with protons roundness in the environment of the threshold energy $E_{Th}$ and the previously assumed condition $Q^{tot}(E)$ for $E < E_{Th}$ does not hold due to the quantum mechanical tunneling effect. This roundness can be amplified by the energy spectrum of the impinging protons resulting from the beam-line and a jump of the fluence decrease of primary protons $\Phi_{pp}$ at $E < E_{Th}$ is prevented (see also Figure 2 with regard to the passage of protons through water). The passage of protons through other media, e.g. calcium, leads to a similar roundness [3]. By that, the quantum mechanical tunneling effect can be best studied by really mono-energetic protons with initial energy $E < E_{Th}$.

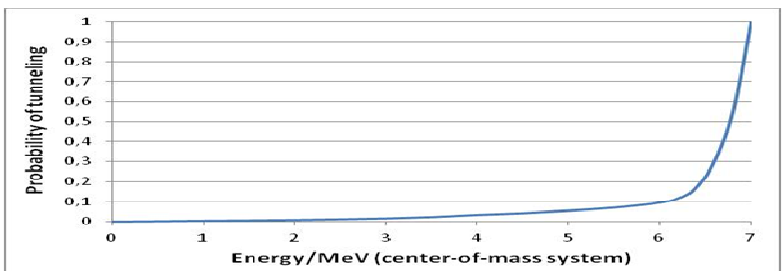

**Figure 8**: Proton – nucleus (oxygen) interaction by proton energies less than 7 MeV via tunnel ling through potential wall

according to Figure 4.

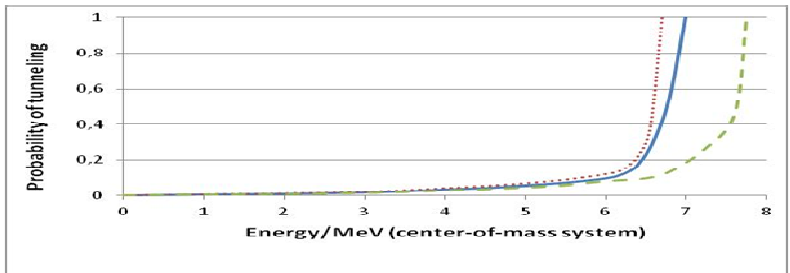

**Figure 9:** Comparison of the tunneling probability for oxygen (solid line), carbon (dots) and calcium (dashes).

Both Figures 8 and 9 also present a way to check ranges of protons in biologically significant media, since at the end of the particle track very specific nuclear interactions occur with the help of the quantum mechanical tunneling effect. The intermediary existence of the isotopes $N_7^{13}$ and $Sc_{21}^{41}$ provides a tool to study proton interactions for very low proton energies in that media. The quantum mechanical tunnel effect represents the necessary basis for these studies.

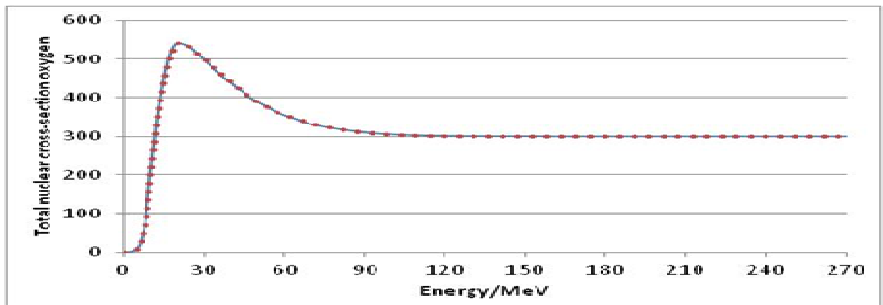

**Figure 10:** Total nuclear cross-section of oxygen with in inclusion of the tunneling effect.

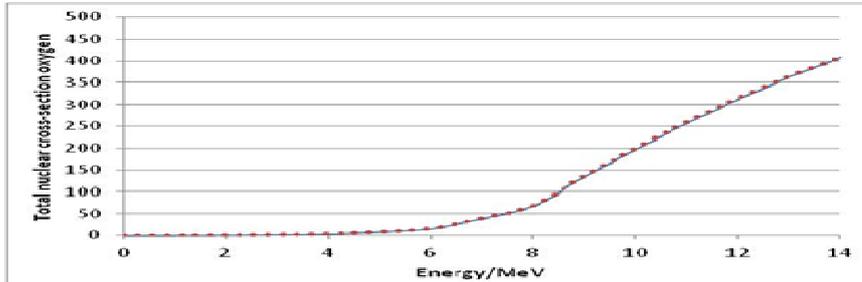

**Figure 11**: See Figure 10 in the low energy domain with proton energy < 14 MeV.

## 4. Discussion

Let us first consider the low energy domain, i.e. up to the resonance maximum $E_{res}$. Thus for $E < E_{Th}$ only due to the tunneling effect the projectile proton can enter the interior of the potential wall according to Figures 9 - 11. This provides in the case of oxygen the isotope $F_9^{17}$ as intermediary excited state (or $N_7^{13}$ and $Sc_{21}^{41}$ with respect to carbon or calcium). Since $E < E_{Th}$ the lifetime of the excited states depends on the transition probability to a lower energy state to finally produce a γ-quant and a neutron by an exchange interaction (Pauli principle) within the nuclei via a $\pi^+$-meson to yield a neutron, which can easily escape the potential wall. The resulting processes are also described by equations (30 – 32).

These equations are also valid for comparably low proton energies, but with $E > E_{Th}$. However, we are now located in the energy domain $E > E_{Th}$ up to the resonance domain $E_{res}$. This domain, which is referred to as Breit-Wigner resonances, is characterized by various excitations, e.g. rotations of the whole nuclei, vibrations via proper deformations of the nuclei and excitations to excited configurations. All these processes are damped by emission of photons with proper energies. It should also be pointed out that these resonances belong to the category of inelastic scatter, but the resulting secondary proton is identical with the primary proton. Only the lateral scatter and energy losses are different to Molière scatter processes. If these scattering processes occur without nuclear excitations, rotations and vibrations, then they are referred to as elastic nuclear scatter of the protons, which are also contained in the Breit-Wigner resonance formula and in its generalization [7], given by Flügge. (The original Breit-Wigner formula [6] is restricted to scatter processes induced by so-called '*s-states*' of the S-matrix; the inclusion of '*p-states*' and states of higher order have later been accounted for [7, 8]). The very often used way of notation 'inelastic nuclear scatter of protons' is not quite correct, since elastic scatter processes of protons at nuclei even occur in the asymptotic domain of $Q^{tot}(E)$. A very important nuclear process occurring in the same energy domain must be mentioned again, namely the exchange interaction of the proton with the mesons of the nucleus, i.e. we consider the reaction:

$$p \Rightarrow \pi^+ \Leftrightarrow (O_8^{16}) \Rightarrow n + F_9^{16} + \gamma. \quad (30)$$

The recoil nucleus $F_9^{16}$ undergoes ß+-decay of electron capture (EC) to become finally again $O_8^{16}$. With regard to nuclear reactions of therapeutic protons the cross-section of oxygen is certainly most important. However, the interaction with calcium (bone) and proteins (carbon) seems also to be worthy of interest. Below we present lists of the most important reactions with carbon and calcium. For brevity, we do not state the decay reactions of the isotopes (detailed listings can be found in web), but for most of them electron capture (EC) and ß+-decay with emission of γ-quanta are preferred reaction channels. For these reasons we have state the symbols X in equations (31 – 31g) and equations (32 – 32h).

As already verified [2 – 3] in the case of oxygen, the nuclear reactions presented in the listings 4.1.1 and 4.1.2 depend on the available or residual proton energy. For this purpose we consider the domain $E > E_{res}$ up to the beginning of the asymptotic behavior, i.e. the proton energy amounts to ca. 100 MeV. We refer to this energy branch of $Q^{tot}(E)$ as $E_{temperate}$. (In the case of copper, the asymptotic behavior is reached at a somewhat higher energy (ca. 150 MeV), but this element is mainly of importance with regard to the determination of beam-line properties). Thus in the energy domain $E_{temperate}$ the nuclear reactions according to relations (31 – 31c) and (32 – 32d) preferably occur, whereas for $E > E_{temperate}$ or $E \gg E_{temperate}$ the release of clusters according to relations (31d – 31g) and (32e – 32h) are more probable. A typical case of a cluster release is the α-particle as a secondary particle. According to [3] this threshold energy amounts to 100 MeV. In the case of carbon it is increased to be 101 MeV, for calcium it amounts to 98 MeV and for copper to 97 MeV. At about E = 190 MeV the probability of this release of α-particles as outcome of nuclear reactions vanishes for all cases considered here. The main difference, however, exists in the yield of this secondary particle. We normalize total yield of α-particles to '1' for the reaction of proton with oxygen. Then in the case of carbon we obtain a total yield of 0.97. For calcium it amounts to 1.68 and for copper to 1.93. Therefore it is generally correct that mainly the yield of clusters is increasing, if the nuclear charge Z and mass number $A_N$ correspondingly increase. A previous check of GEANT4 [10] in the papers [1 – 3, 13, 15] revealed that this Monte-Carlo code provides default reaction channels, which underrate cluster formations. However, since this Monte-Carlo code is an open system, it can be improved by the user. A further lack of this Monte-Carlo code is that the release of neutrons is not sufficiently accounted for, as we could verify at a glance at the papers [11, 12]. According to the present results the release of low energy neutrons should be much more in focus in dose calculations due to the rather high RBE.

The calculation method for the total nuclear cross-section $Q^{tot}(E)$ provides a significant advantage compared to the previously published method [2, 3]: Each Gaussian distribution and the error-function distribution with a specific energy shift can be associated to probability distributions for the occurrence of some nuclear reaction types. Thus the energy shifts now refer to threshold energies with regard to the corresponding maxima and the half-breadths represent a measure for the yield of some reaction types. Furthermore, it is possible to analyze measured curves of $Q^{tot}(E)$ by suitable deconvolution procedures as worked out in [16]. These convolution methods are developed with regard to linear combinations of Gaussian kernels and shifts. Since there exists a connection between the number of kernels and the underlying statistics, which is strictly a non-relativistic Boltzmann distribution in the case of one Gaussian kernel and a linear combination of different Gaussian kernels with further shifts in the case of the Fermi-Dirac statistics, we have also obtained a way to interpret nuclear reactions by the operator formulation of Fermi-Dirac statistics [2, 16]. Such statistical models can help to calculate nuclear interaction processes in a simpler way compared to many-body-problems of relativistic quantum mechanics. The evaluation of measurement data in clinical practice of proton therapy, e.g. the contributions of secondary particles, requires relatively simple but reliable methods.

### 4.1. Listings of possible nuclear reactions of some nuclei of particular interest

Nuclear reaction types of protons with carbon and calcium (with oxygen already reported in [1 - 3]).

### 4.1.1. Interactions proton – carbon:

$$p + C_6^{12} \Rightarrow n + N_7^{12} + X . \quad (31) , \quad p + C_6^{12} \Rightarrow n + p + C_6^{11} + X . \quad (31\,a)$$

$$p + C_6^{12} \Rightarrow n + 2p + B_5^{10} + X . \quad (31\,b) , \quad p + C_6^{12} \Rightarrow 2n + 2p + B_6^{9} + X . \quad (31\,c)$$

$$p + C_6^{12} \Rightarrow H_1^2 + C_6^{11} + X. \quad (31d), \quad p + C_6^{12} \Rightarrow H_1^3 + C_6^{10} + X. \quad (31e)$$

$$p + C_6^{12} \Rightarrow He_2^3 + B_5^{10} + X. \quad (31f), \quad p + C_6^{12} \Rightarrow He_2^4 + B_5^9 + X. \quad (31g)$$

**4.1.2. Interaction proton – calcium:**

$$p + Ca_{20}^{40} \Rightarrow n + Sc_{21}^{40} + X. \quad (32), \quad p + Ca_{20}^{40} \Rightarrow n + p + Ca_{20}^{39} X. \quad (32a)$$

$$p + Ca_{20}^{40} \Rightarrow 2n + p + Ca_{20}^{38} + X. \quad (32b), \quad p + Ca_{20}^{40} \Rightarrow 2n + 2p + K_{19}^{37} + X. \quad (32c)$$

$$p + Ca_{20}^{40} \Rightarrow n + 2p + K_{19}^{38} + X. \quad (32d), \quad p + Ca_{20}^{40} \Rightarrow H_1^2 + Ca_{20}^{39} + X. \quad (32e)$$

$$p + Ca_{20}^{40} \Rightarrow H_1^3 + Ca_{20}^{38} + X. \quad (32f), \quad p + Ca_{20}^{40} \Rightarrow He_2^3 + K_{19}^{38} + X. \quad (32g)$$

$$p + Ca_{20}^{40} \Rightarrow He_2^4 + K_{19}^{37} + X. \quad (32h)$$

Please note that X stands for further nuclear decay reactions, which preferably incorporate ß[+], ß[-] decay and electron capture (EC). In particular, the emission of positrons is associated for further radiation processes by pair annihilations.